# Shoot-through layers in upright proton arcs unlock advantages in plan quality and range verification

Erik Engwall, Victor Mikhalev, Johan Sundström, Otte Marthin, Viktor Wase
*RaySearch Laboratories, Stockholm, Sweden*

**Abstract**

*Background:* Upright proton therapy with compact delivery systems has the potential to reduce costs for treatments but could also lead to broadening of the beam penumbra due to energy selection close to the patient.

Purpose: This study aims at combining upright static proton arcs with additional layers of shoot-through (ST) protons to sharpen the beam penumbra and improve plan quality for such systems. An additional advantage of the method is that it provides a straightforward approach for range verification with a fixed range detector opposite the fixed proton nozzle.

*Methods:* We examined various treatment plans for a virtual phantom: 3-beam IMPT, static arc (Arc) with/without ST (Arc+ST), and with/without collimation (+Coll). In the virtual phantom three different targets were utilized to study the effect on conformity index (CI), homogeneity index (HI), robustness and mean dose to the phantom volume. The phantom study was complemented with a head-and-neck (H&N) patient case with a similar set of plans. The delivery time for all plans was estimated using a combined model of the upright patient positioner and the proton nozzle. A range verification concept that determines residual ranges of the ST protons was studied in simulated scenarios for the H&N case.

*Results:* In the phantom study, the Arc+ST plans show superior CI, HI and target robustness compared to the Arc+Coll plans. For the Arc plans without ST, the collimated plans perform better than the uncollimated plans. On the other hand, for Arc+ST, collimation has little impact on CI, HI and robustness. However, a small increase in the mean dose to the phantom volume is seen without collimation. For the H&N case, similar improvements for Arc+ST can be seen with only a marginal increase of the mean dose to the patient volume when no collimation is used. These results imply that no aperture is needed when combining arcs with ST, which in turn substantially reduces treatment times: for the H&N case the delivery time for Arc+ST is estimated to 5.4 minutes and for Arc+Coll to 6.5 minutes. The range verification simulation shows that the method is sensitive to detect systematic stopping power ratio errors, setup errors and changes in the patient anatomy.

*Conclusions:* Combining proton arcs and ST layers can enhance compact upright proton solutions by improving plan quality at the same time as delivery time is reduced. It is also tailored for the inclusion of a fast and straightforward residual range verification method.

## 1 Introduction

During recent years, interest in proton arc therapy (PAT) has grown significantly, due to its potential to improve plan quality and reduce treatment times[1]. Numerous treatment planning studies have shown dosimetric benefits of PAT over conventional intensity modulated proton therapy (IMPT) across various treatment sites[2–12]. PAT can be categorized into two types: *dynamic* and *static arcs*. Dynamic arcs are delivered while the patient (or gantry) is rotated, e.g., spot scanning and energy switching occur during rotation, enabling a fast delivery. Static arcs, on the other hand, involve delivery of multiple energy layers from a number of discrete angles (typically 20–30) using a step-and-shoot delivery approach[1,2,12,13]. Static arcs have the advantage over dynamic arcs of not introducing any potential dosimetric uncertainties beyond those of IMPT, since the beam or patient remains stationary during beam delivery. Furthermore, static arcs can be delivered on any conventional proton machine by converting the arc into a standard IMPT plan[13,14].

The first clinical PAT treatments using static arcs were reported by Fracchiolla et al.[12] following a comprehensive study of 10 patient cases, comparing plan quality of PAT to state-of-the-art clinically delivered IMPT plans (non-coplanar beam directions and range shifter splitting technique). Their results align with those reported on static arcs by de Jong et al. in a number of studies[2–4]. These studies consistently showed improvements in plan quality while simplifying beam arrangements through a coplanar arc setup without range shifters.

Such simplifications enabled by proton arcs facilitate the transition to treatments with a fixed horizontal beamline and rotation of the patient in an upright position – a setup which has more limited degrees of freedom compared to a traditional gantry-based facility, where both the gantry and the treatment couch can be rotated. While an upright patient positioner can be combined with any existing horizontal fixed beam line to increase the accessible treatment angles, turn-key compact solutions are being developed[1,15]. The Mevion s250-FIT system (Mevion Medical Systems, MA, USA), for example, combines a compact proton cyclotron with an upright positioner and an in-room CT scanner for retrofitting in existing photon treatment bunkers[1]. However, this compactness comes at the expense of energy selection near the patient, creating more scattering in the proton beam, an issue partly mitigated by the use of an adaptive aperture to sharpen the penumbra of each spot[16].

Kong et al.[17] studied the impact of adding shoot-through (or transmission) layers of protons to conventional IMPT plans with a few beam directions and saw considerable reductions in doses to organs at risk (OARs). This idea aligns with approaches to combine photons with protons to exploit the sharp penumbra of photons, while keeping the targeting effect of protons[18]. We hypothesize that shoot-through layers could be even more beneficial when combined with proton arcs due to the many beam directions, reducing the dose level delivered by the shoot-through protons behind the target. We therefore propose incorporating a shoot-through (ST) layer for each direction in a static arc plan to investigate potential dosimetric advantages of this combination. The combination of ST layers and proton arcs has a general applicability, and it could be employed by most delivery machines, including gantry-based facilities. However, it is particularly suited for upright treatments, where the protons will exit the patient in the same direction for all treatment angles, allowing for reduced shielding of the shoot-through protons using a single beam dump opposite the fixed beam. Additionally, equipping the beam dump with a fixed range detector could enable almost instantaneous and straightforward range verification of proton beams.

Accurate range verification in proton therapy is highly desirable due to the protons' high sensitivity to uncertainties. Several approaches for in-vivo range verification have been explored in the literature[19]. Secondary particle detection, such as prompt gamma detection[20,21] and PET-based range verification[22,23], has been studied extensively, but so far, the systems have not reached any full-scale clinical implementation due to the complexity of the reconstruction and detection methods[19]. The most pragmatic approach is to measure the residual range of protons exiting the patient by detectors suitable for proton radiography[24,25]. This method has been implemented in a clinical setting to routinely determine range errors in patients[26]. It has also been used to validate the accuracy of synthetic CTs from daily CBCTs.[27]

In this paper, we report on a proof-of-concept study for the combination of static proton arcs, ST layers and upright position for three different target geometries in a virtual phantom, as well as for a head-and-neck (H&N) patient case. Furthermore, we investigate the feasibility of detecting range errors by measuring the residual proton range of the ST layers.

# 2 Methods

## 2.1 General concept

The general concept of incorporation of ST layers in upright proton arcs is illustrated in Figure 1. Bragg peaks in the target are combined with ST protons to utilize the sharp penumbra of the high-energy protons. The optimizer will determine where to place the ST spots based on the dose-based objectives, as described in section 2.2.2. The shoot-through protons, exiting in the same direction behind the patient, could be used for near-instantaneous range probing during treatment from a large number of directions. The range verification method is explained in section 2.7.

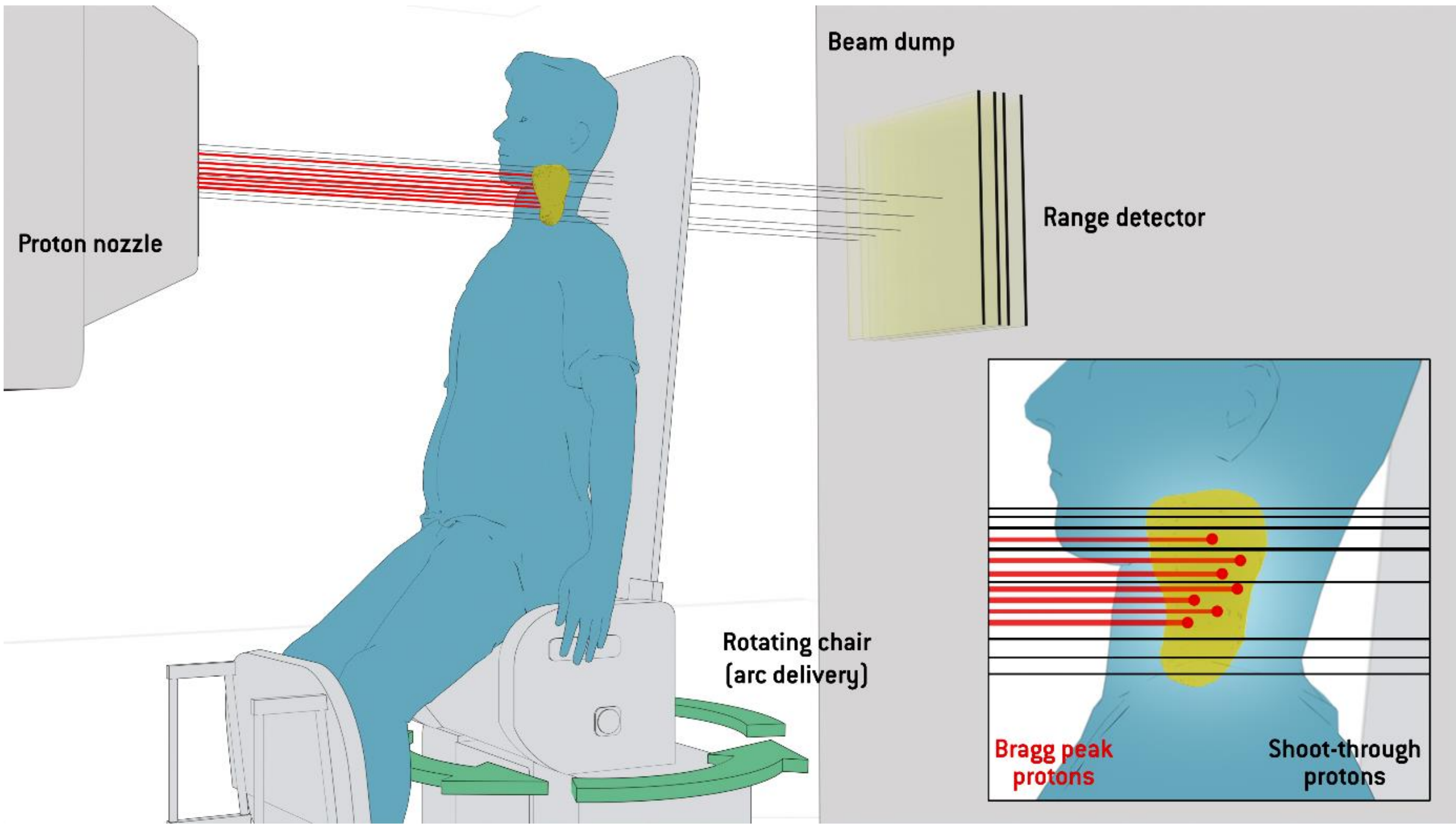

***Figure 1.*** *Schematic sketch of the concept for upright static arc delivery with a shoot-through portion. The patient is placed in an upright patient position and the patient is rotated to the planned treatment angles. The beam is turned off during rotation and is only delivered at discrete directions. The majority of protons from each direction is delivered with the Bragg peaks placed in the target, while a smaller portion is delivered by shooting through the target (including margins) by utilizing the highest energy of the machine. A static range detection device can be integrated into the beam dump behind the patient, in order to verify the range of the protons.*

## 2.2 Treatment plan generation

### 2.2.1 General planning

Both IMPT and PAT treatment plans were generated in a development version of RayStation v2025 (RaySearch Laboratories, Stockholm, Sweden) with a beam model of the Mevion s250-FIT system, which employs binary range shifter plates for energy selection in the nozzle resulting in nominal proton energies from 5 to 230 MeV. Due to the energy selection in the nozzle, the spot delivery current is the same for all energies. The PAT plans were planned in static mode, where the user selects the desired number of directions, and a number of initial energy layers which are filtered in the subsequent optimization to reach a final number of desired energy layers.[13,14] All arc plans in this study had 20 uniformly spaced directions each. For the collimated plans, a layer-by-layer collimation was employed to trim the spots at the boundary of the field for each energy layer[16]. The minimum spot meterset in the beam model was 0.0683 MU, corresponding to 5.63 million protons, and this value was used in the optimization to filter out spots below this threshold. All plans in the study were optimized and computed with the RayStation Monte Carlo dose engine[13]. A constant RBE of 1.1 was employed.

### 2.2.2 Setup and optimization of ST layers

The shoot-through (ST) plans were set up with an additional energy layer of the highest energy (230 MeV) for each direction, both for the IMPT and PAT plans. The spot selection for the ST layer fills the full target projection (including margins) with spots. In the subsequent optimization, the objectives in combination with the spot filtering will determine where the ST spots will remain and have high weights.

## 2.3 Phantom plans

### 2.3.1 Treatment geometry

To facilitate the study of different target sizes and make the study more reproducible, a virtual phantom was used as the main geometry in this study (see Figure 2). The outline (External ROI) is an ellipsoid (length, $l$=20 cm, height, $h$=20 cm, width, $w$=17 cm) with an override of water. Inscribed in the ellipsoid were an air cylinder (diameter, $d$=2 cm, $l$=5cm), a bone cylinder ($d$=3.5 cm, $l$=8 cm), as well as an ellipsoid ($l$=4 cm, $w$=6 cm, $h$=8 cm) representing an OAR. Three different targets were placed in the center of the External and were used separately in different treatment plans:

- CTV Small: cube with sides of 3 cm
- CTV 4.5H: cube with sides of 4.5 cm and two cutout cuboids ($l$=1.5 cm, $w$= 1.5 cm, $h$=4.5 cm), forming an H-shape.
- CTV 6H: cube with sides of 6 cm with two cutout cuboids ($l$=2 cm, $w$=2 cm, $h$=6 cm).

The plans were planned with 30 fractions each. The same objective functions were used for all plans: the CTVs had a uniform dose objective of 60 GyRBE (weight 20), as well as min DVH (weight 5) and max DVH (weight 10) objectives at 60 GyRBE to 95% and 2% volume, respectively. The OAR had a max EUD objective (parameter A=1) of 4 GyRBE (weight 3). The objectives for the target and the OAR were defined to be robust to 3 mm setup shifts and 3.5% range errors. For the External, two dose fall-off functions were employed: from 60 GyRBE to 0 GyRBE over 1 cm (nominal weight 0.5) and from 60 GyRBE to 10 GyRBE over 0.5 cm (nominal weight 1). The dose fall-off weights were changed from the nominal weights in variations of the same plans, as is described in the next section.

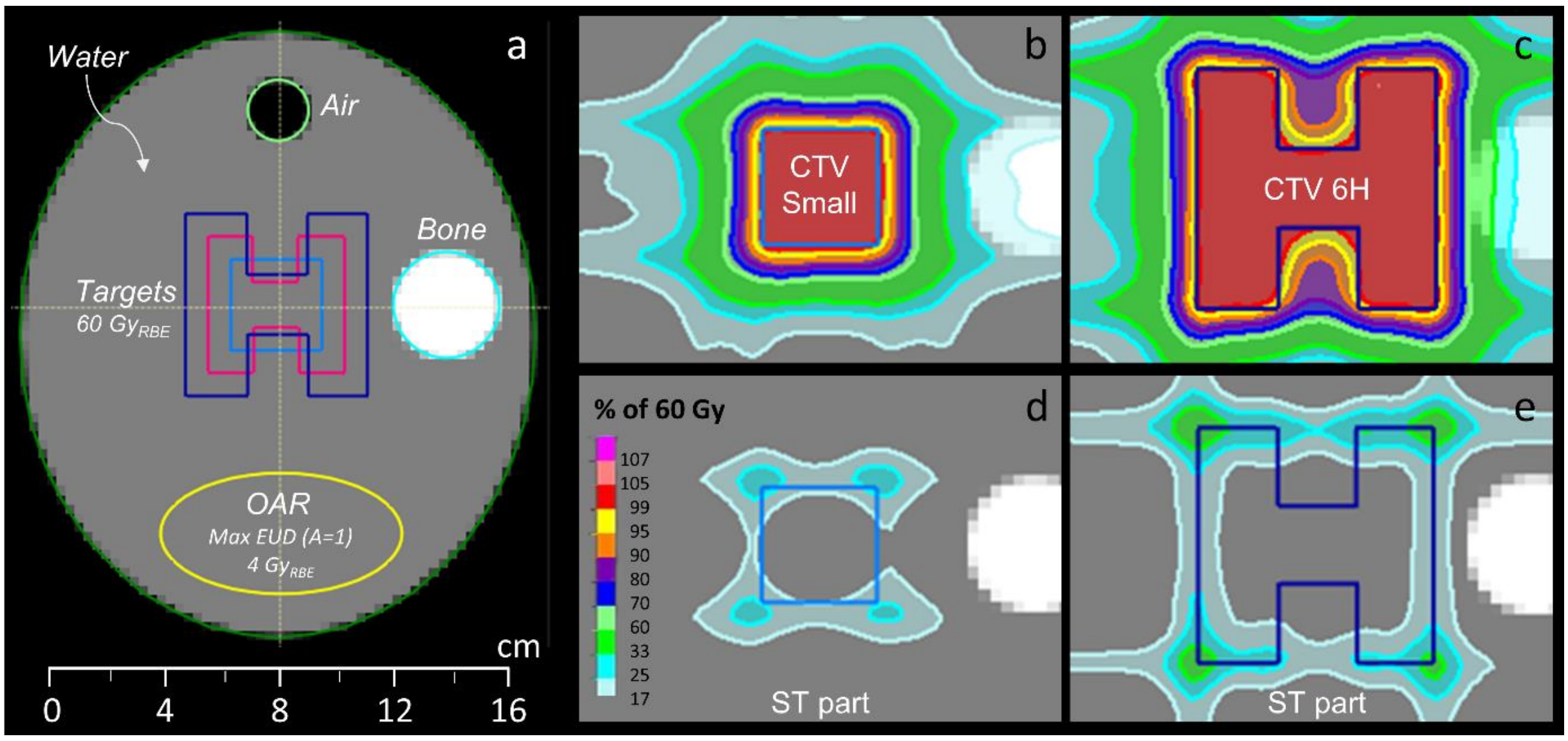

***Figure 2.*** *(a) The simulated geometry with the three targets: CTV 6H (dark blue), CTV 4.5H (magenta) and CTV Small (light blue). The main volume consists of water but also includes an air cylinder with diameter 2 cm and length 5 cm (light green) and a bone cylinder with diameter 3.5 cm and length 8 cm (cyan). The yellow ellipsoid simulates an OAR with a max EUD objective (A=1) of 4 GyRBE. (b)-(c): Plan dose for two of the three artificial*

*targets for the Arc+ST plan: CTV Small and CTV 6H. The prescription to the targets is set to 60 GyRBE. (d)-(e): Part of the dose distribution delivered by shoot-through protons.*

### 2.3.2 Set of plans

For each target, a large set of plans was generated:

- Static arc plans without ST (Arc)
- Static arc plans with ST (Arc+ST)
- 3-beam IMPT plans without ST (3-beam)
- 3-beam IMPT plans with ST (3-beam+ST)

Each of the arc plans were generated with different layer-by-layer collimation: no collimation and collimation (+Coll). For the 3-beam plans, a collimator was always used to increase conformity. The lateral target margins were based on the spot size multiplied by a factor[13]: for the uncollimated plans the factor was always set to 1.0 and for the collimated plans factors of 1.0, 1.25 and 1.5 were explored. Additionally, for the arc plans the dose fall-off weight was increased from the nominal weight by factors of 2, 3 and 4. The initial and final number of energy layers in the static arc optimization were 480 and 240, respectively.

## 2.4 Patient plan

### 2.4.1 Treatment geometry

In addition to the main part of the study with a large set of plans for the three target geometries in the virtual phantom, one H&N patient case (laryngeal cancer) was investigated. In this case, the dose was planned for 33 fractions with a 66 GyRBE prescription to the primary target (CTV66) and a 55 GyRBE prescription to the secondary target (CTV55). Optimization objectives were defined for the targets, as well as for the spinal cord (max dose of 48 GyRBE and max EUD (A=15) of 15 GyRBE), the parotids (max EUD (A=1) of 10 GyRBE) and the submandibular glands (max EUD (A=1) of 10 GyRBE). Robust optimization objectives with 3 mm setup and 3.5% range uncertainties were defined for the targets, as well as for the max dose to the spinal cord.

A 2D view of the patient geometry with beam setup, doses and DVHs from three of the plans can be seen in Figure 4.

### 2.4.2 Set of plans

A similar set of plans as for the virtual phantom was created for the H&N case, but with the following modifications: (1) 3-beam plans were replaced by 5-beam plans, which are more commonly used in a clinical setting for H&N patients, (2) for plans with a collimator, the target margin was set to 1.25 for all cases, since this was the best performing target margin in the virtual phantom, and (3) the full 360-degree arc plans were complemented with sub-arc plans, consisting of two sub-arcs and two lateral beams at 90° and 270°, in order to avoid that the protruding shoulders enforce a large air gap over the full revolution. For the patient case, the initial and final number of energy layers were 720 and 360, respectively, which is in accordance with settings from previous studies with H&N patients[2,12].

## 2.5 Plan quality evaluation metrics

All plans were evaluated with respect to:

- Paddick conformity index (CI)[28].
- Homogeneity index (HI), defined as D95/D5 for the CTV.
- The mean dose to the External ROI.
- The worst-case scenario for the CTV D95 in a robustness evaluation with 28 scenarios using 3.5% density and 3 mm setup uncertainty.

For the H&N case, doses to OARs (parotids, submandibular glands and spinal cord) were also investigated. In addition to the explicit plan quality evaluation metrics, we report the number of energy layers and spots, as well as the portion of ST delivered.

## 2.6 Delivery time estimation

We have employed a comprehensive model of the upright patient positioner and the proton nozzle. The rotational speed and acceleration/deceleration have been estimated to 6°/s and 5°/s$^2$, respectively. In the model of the proton delivery system, the average scan speed is set to 600 mm/s and the spot delivery current to 55 nC/s. Energy switching time is estimated to be around 0.2 s for a centrally located spot. The leaf movement of the adaptive aperture is modeled at a speed of 500 mm/s.

## 2.7 Range verification

For range verification we follow the method described by Mumot et al.[29] and clinically applied in Meijers et al.[26] and Seller Oria et al.[27], where the residual range for each ST spot is first simulated on the planning CT geometry by the TPS and then compared to the measured range to assess the magnitude of residual range deviations. Since we do not have access to measurement equipment or real patient data, we use the TPS to simulate the measurements on a perturbed CT geometry, serving as a conceptual demonstration of the measurement-based range verification method.

We compute the residual range by a ray-trace through the patient geometry including the material composition of each voxel. To investigate the sensitivity of the range verification concept to detect water-equivalent thickness (WET) differences in the patient, we have performed simulations for three different error scenarios: 5% systematic shift in the stopping power ratio (SPR) of the patient volume, 5 mm setup shift in the anterior direction, and 2 cm movement of the right shoulder.

We have used the residual ranges, $r$, from the original treatment plan and from the simulated range measurement to compute the relative difference in WET, $\Delta WET/WET$, for each individual ST spot:

$$\frac{\Delta WET}{WET} = \frac{(R_{\max} - r_{\text{measured}}) - (R_{\max} - r_{\text{plan}})}{R_{\max} - r_{\text{plan}}} = \frac{r_{\text{plan}} - r_{\text{measured}}}{R_{\max} - r_{\text{plan}}}$$

where $R_{\max}$ is the proton range in water of the ST protons and is a known intrinsic property of the machine, and $r_{\text{measured}}$ and $r_{\text{plan}}$ are the residual ranges from (simulated) measurements and from computation based on the treatment plan, respectively. This means that in order to assess the relative WET difference in a real treatment situation, two steps are needed: (1) measurement of the residual range behind the patient using a range detector, and (2) estimation of the residual range by tracing the protons from the nominal plan through the CT geometry. If notable deviations are found between the measurement and estimation, the therapists can be notified that the delivered dose might differ from the planned dose.

# 3 Results

## 3.1 Plan quality metrics and delivery time

### 3.1.1 Phantom plans

In Figure 3, CI is plotted for the CTV 6H target as a function of three different quantities: HI, worst-case D95 of the CTV over all robustness scenarios, and the mean dose to the External. Each plan type is represented by four data points, representing the different fall-off objective weights employed. The best performing collimated arcs were the ones with target margin of 1.25 times the spot size and those plans are used in the subsequent analysis. For the CI-HI relationship, the Arc+ST plans follow a Pareto front, which shows better performance than the arc plans without ST. The same trend can be seen for CI versus robust target coverage. The 3-beam plans (not shown in Figure 3) achieve similar HI and similar or better robustness compared to the best arc plans, but with a considerably lower CI (see Table 1). Uncollimated arc plans without ST show worse performance than their collimated counterparts in terms of HI and robustness, whereas no notable difference can be seen between collimated and uncollimated arc plans with ST. For the mean dose to External, all uncollimated arc plans result in more dose than the collimated plans. The plans with increased dose fall-off weights improve, as expected, the CI at the expense of reduced HI and robustness.

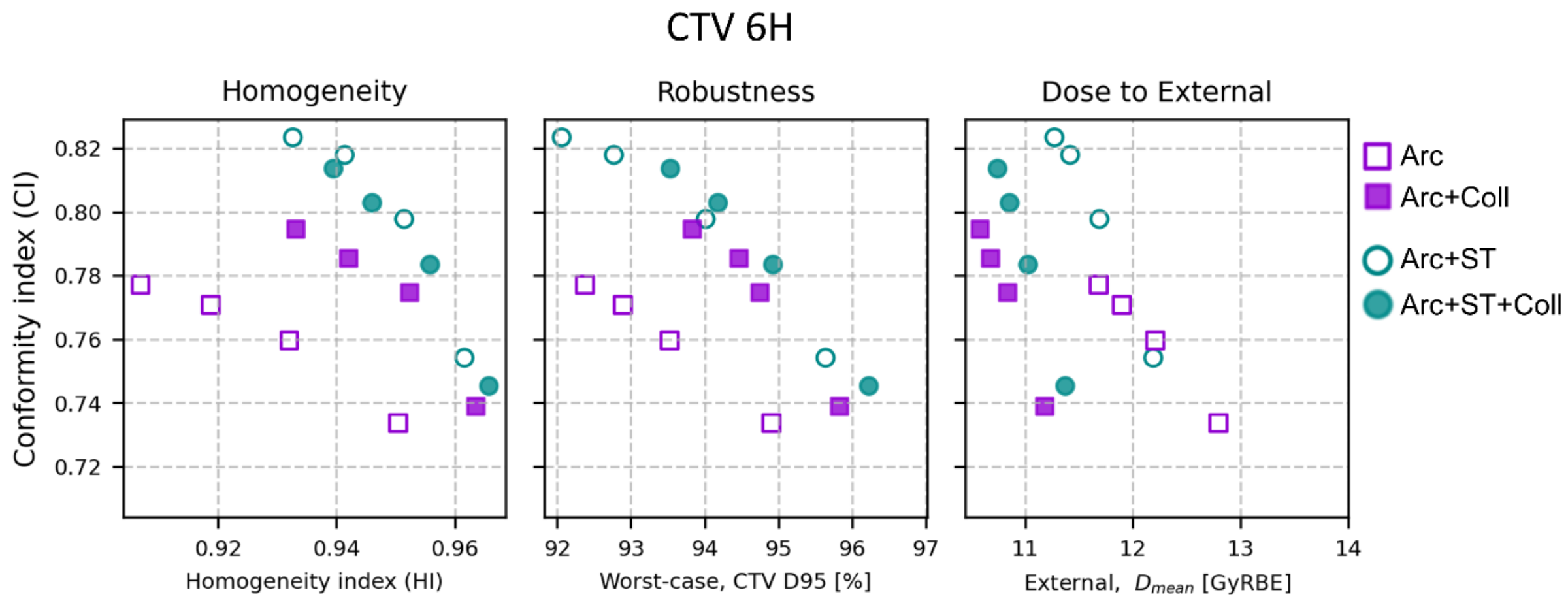

***Figure 3.*** *Evaluation for the CTV 6H: Paddick conformity index, CI, as a function of homogeneity index, HI, (left), worst-case robustness scenario for the CTV D95 (given in percentage of the prescription dose; middle), and mean dose to the External ROI (right). Collimated plans are shown with filled symbols, while uncollimated plans are shown with open symbols. The target margin for collimated (uncollimated) plans is 1.25 (1.0) times the spot size. The plans have different weights for the dose fall-off objective (factors 1, 2, 3, 4 compared to the nominal weights). The higher the weight, the higher conformity is in general achieved, and those plans reside in the upper left part of the plots.*

The same trends as for CTV 6H are seen for the other two targets (see Figure S1; supplementary material). In Table 1, metrics are summarized for the Arc+ST, Arc+ST+Coll, Arc+Coll, 3-beam and 3-beam+ST plans for the three targets. The plans in the table all used nominal dose fall-off weights. It can be noted that for the 3-beam plan, there is no clear gain for the conformity by adding the ST portion. While the ST protons will contribute to a sharper penumbra, they will also contribute to a higher dose behind the target. In contrast to the static arcs, the 3-beam plans cannot distribute this exit dose over many directions and therefore the conformity as a whole might not be improved with ST protons.

| | Plan | No. EL | No. spots | % ST | Protons ($10^9$) | Delivery time [s] | CI | HI | Worst-case CTV D95 | External, $D_{mean}$ [$Gy_{RBE}$] |
|---|---|---|---|---|---|---|---|---|---|---|
| CTV Small | *Arc+ST* | *234* | *1939* | *30.8* | *44* | *183* | *0.73* | *0.97* | *97.3%* | *4.79* |
| | Arc+ST+Coll | 239 | 2714 | 19.1 | 59 | 231 | 0.72 | 0.97 | 97.2% | **3.94** |
| | Arc+Coll | 240 | 1803 | 0 | 52 | 254 | 0.72 | 0.95 | 95.8% | **3.79** |
| | 3-beam | 37 | 433 | 0 | 57 | 75 | 0.72 | 0.94 | 95.7% | **3.76** |
| | 3 beam+ST | 34 | 627 | 20.7 | 53 | 77 | 0.68 | 0.96 | **98.3%** | **4.10** |
| CTV 4.5H | *Arc+ST* | *240* | *4033* | *25* | *75* | *260* | *0.59* | *0.97* | *96.4%* | *8.13* |
| | Arc+ST+Coll | 240 | 5432 | 14.6 | 98 | 427 | 0.58 | 0.97 | **96.9%** | **7.11** |
| | Arc+Coll | 240 | 4849 | 0 | 87 | 432 | 0.58 | 0.96 | 96.1% | **7.03** |
| | 3-beam | 41 | 957 | 0 | 99 | 110 | 0.58 | 0.95 | 96.1% | **7.16** |
| | 3-beam+ST | 39 | 1111 | 12.2 | 92 | 115 | 0.57 | 0.96 | **97.5%** | **7.38** |
| CTV 6H | *Arc+ST* | *240* | *6582* | *24.1* | *111* | *361* | *0.75* | *0.96* | *95.6%* | *12.19* |
| | Arc+ST+Coll | 240 | 7951 | 13.8 | 139 | 598 | 0.75 | **0.97** | **96.2%** | **11.37** |
| | Arc+Coll | 240 | 7493 | 0 | 126 | 587 | 0.74 | 0.96 | 95.8% | **11.18** |
| | 3-beam | 47 | 1624 | 0 | 142 | 159 | 0.71 | 0.96 | **96.8%** | **11.71** |
| | 3-beam+ST | 47 | 1621 | 7.4 | 136 | 156 | 0.73 | 0.96 | **96.2%** | **11.85** |

***Table 1.*** *Plan metrics for the artificial targets CTV Small, CTV 4.5H and CTV 6H. Three different arc plans are compared for each CTV: (1) arc with additional shoot-through layers (Arc+ST), (2) collimated arc with additional shoot-through layers (Arc+ST+Coll), and (3) collimated arc (Arc+Coll). In addition to the arc plans, collimated 3-beam plans are shown in the table, both with and without additional shoot-through layers. Number of energy layers and spots, as well as the percentage of shoot-through protons (% ST) and the total number of irradiated protons are shown in the first four columns. The estimated delivery time is reported in column 5. Paddick conformity index (CI) is reported, along with homogeneity index (HI), defined as D95/D5. Robust evaluation was performed over 28 scenarios (3 mm setup/3.5% density uncertainty) and the worst-case value in percent (relative the prescription dose 60 GyRBE) of the CTV D95 is reported. The last column gives the mean dose to the External ROI. A bold value signals that the value is better than in the corresponding Arc+ST plan.*

The portion of ST is higher for the uncollimated arcs than for the collimated arcs and ranges between 24.1% to 30.8% of the total number of irradiated protons. In general, objectives that aim for a sharp dose fall-off in the vicinity of the target generally favor a greater portion of ST protons. Conversely, objectives that aim for low dose to the entire patient volume will suppress the portion of ST protons. As can be seen in Figure 2, the ST protons are placed at the edges of the target to sharpen the penumbra. It can be noted that there is not a large portion of ST protons placed at the inner edges of the H-structures. This can be explained by the fact that the openings are 1.5 cm and 2 cm for the CTV4.5H and CTV6H, respectively, which is small compared to the dose fall-off distance of 1 cm. This is especially pronounced for the CTV4.5H, for which the conformity is substantially lower than for the other targets.

The number of energy layers and spots are substantially higher for the arc plans compared to the IMPT plans, but they are comparable between the different arc plans. It can be noted that the final number of energy layers in the static arc plans can be slightly lower than requested by the user due to removal of complete energy layers in the spot filtering step. The estimated delivery times for the Arc+ST plans are shorter than the collimated arc plans, which can mainly be attributed to the removal of the time for the movement of the adaptive aperture. Furthermore, the total number of protons used in the irradiation for Arc+ST is substantially lower than for the plans with aperture. This is expected since a portion of the protons will be absorbed by the aperture in collimated plans. Compared to 3-beam plans, the arc plans exhibit a longer delivery time but in terms of irradiated protons the Arc+ST plans deliver the lowest numbers.

### 3.1.2 Patient plans

For the H&N case, both arc plans with ST (Arc+ST and Arc+ST+Coll) perform better than all other plans for all metrics, except for a small increase in the mean dose to the External (see Table 2). However, the increase in mean dose is much smaller than for the phantom cases. The Arc+ST+Coll plan shows some marginal improvements for some of the metrics compared to the Arc+ST, e.g., mean dose to External, but in essence the two plans are of similar plan quality. As for the phantom plans, the ST protons are primarily placed where a sharp penumbra is desirable, e.g., at the border of the CTV (see Figure 4), and especially in regions where the CTV is close together to OARs. The latter effect can clearly be seen in Figure S3 in the supplementary material, where a large portion of ST protons are delivered at the border between the CTV 66 and the submandibular glands.

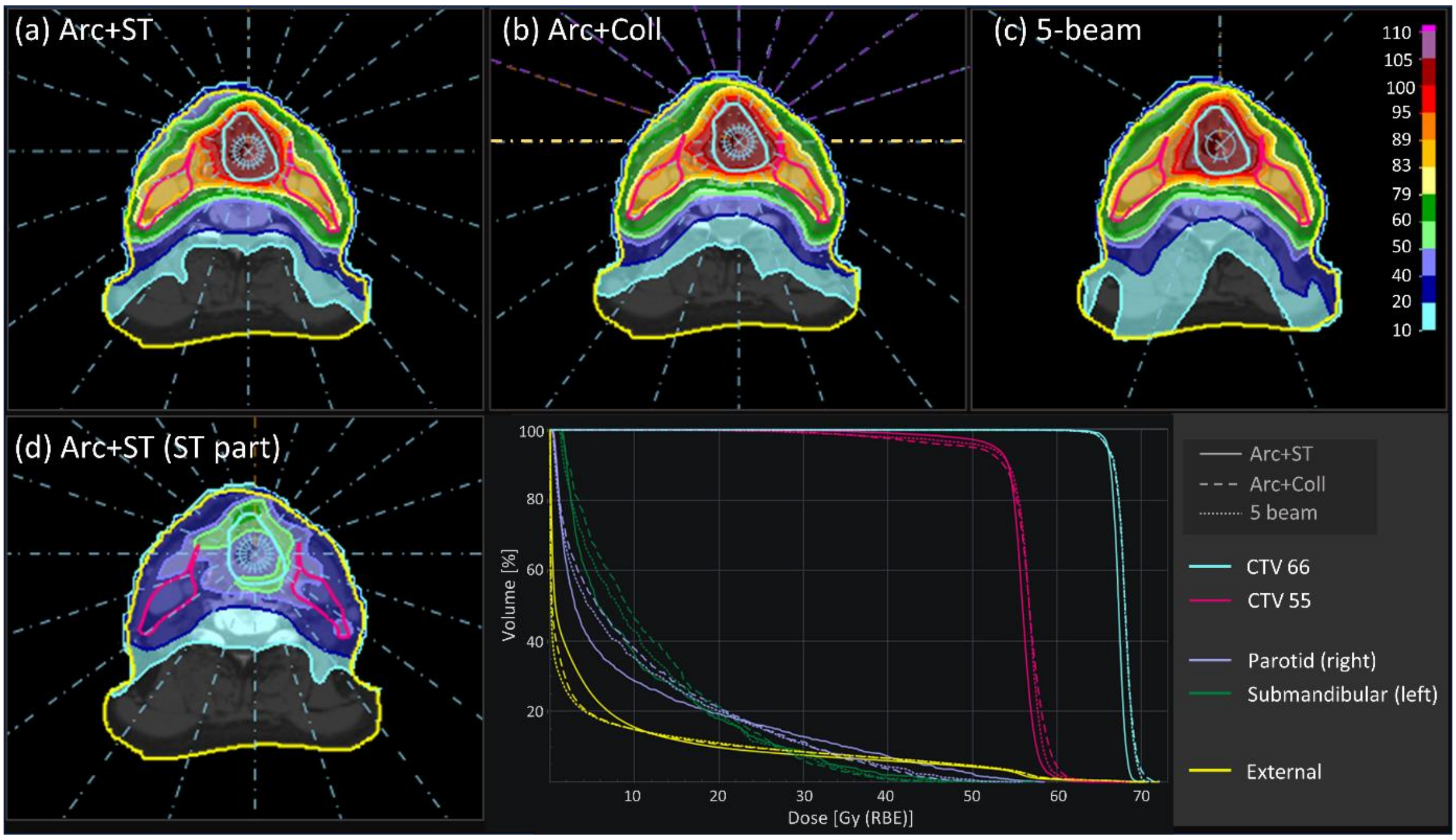

***Figure 4.*** *Comparison of different planning techniques for the H&N case: (a) Static arc with an additional ST layer (Arc+ST), (b) Partial static arcs with collimation (sub-Arc+Coll), and (c) 5-beam collimated plan (5-beam). In panel (d) the ST portion of the dose distribution in (a) is displayed. In the lower right, the DVHs for the doses in (a), (b) and (c) are displayed for the targets and representative OARs. (Statistics for more OARs can be found in Table 2 and enlargement of the target DVHs are supplied in Figure S4 in the supplementary material.) The primary target prescription is 66 GyRBE and the secondary target prescription is 55 GyRBE. The color scale in panels (a)–(d) represents percentage of the primary target prescription of 66 GyRBE. The legend for the colors for the targets and the OARs applies both to the patient views and the DVHs. (A coronal view of the Arc+ST plan is given in Figure S3 in the supplementary material.)*

It can be noted that the sub-Arc+Coll shows a better HI and lower mean dose to External than the Arc+Coll, which can be explained by the influence of the air gap on the plan quality. For the Arc+ST plan, on the other hand, the effect of splitting up the full arc into sub-arcs is minor (see Table S1;

supplementary material). The 5-beam plan performs at the same level as the sub-Arc+Coll plan, (see Table 2 and Figure 4 ). Also for the 5-beam plan, the plan quality metrics are improved when adding an extra ST layer per beam direction, which is in line with the findings from Kong et al.[17] The portion of ST is lower for the IMPT plan (around 16% for 5-beam) compared to the Arc+ST plan (45%).

| | H&N case | | | | | |
|---|---|---|---|---|---|---|
| | *Arc+ST* | Arc+ST+Coll | Arc+Coll | sub-Arc+ +Coll | 5-beam | 5-beam+ST |
| No. EL | *360* | 340 | 359 | 358 | 162 | 154 |
| No. spots | *4281* | 3594 | 2255 | 3003 | 2707 | 3077 |
| %ST | *44.8* | 37.2 | 0.0 | 0.0 | 0.0 | 15.7 |
| Protons ($10^9$) | *223* | 262 | 377 | 371 | 327 | 292 |
| Delivery time [s] | *321* | 425 | 387 | 452 | 342 | 357 |
| CI, CTV66 | *0.48* | 0.48 | 0.42 | 0.41 | 0.41 | 0.43 |
| CI, CTV55 | *0.36* | **0.37** | 0.32 | 0.33 | 0.33 | 0.33 |
| HI, CTV66 | *0.96* | 0.96 | 0.93 | 0.95 | 0.95 | 0.96 |
| HI, CTV55 | *0.91* | 0.91 | 0.82 | 0.84 | 0.88 | 0.89 |
| Worst-case, CTV66 D95 | *99.2%* | **99.3%** | 99.2% | 99.1% | 98.7% | 98.8% |
| Worst-case, CTV55 D95 | *76.5%* | **76.9%** | 70.9% | 72.1% | 72.8% | 74.2% |
| OAR dose [$Gy_{RBE}$] | | | | | | |
| External, $D_{mean}$ | *6.4* | **6.1** | 6.7 | **6.1** | **6.0** | **6.0** |
| SpinalCord, $D_{max}$ | *20.1* | 23.0 | 23.9 | 20.6 | 22.3 | 25.4 |
| Parotids (L+R), $D_{mean}$ | *10.3* | **10.2** | 10.6 | 10.7 | 10.6 | 10.5 |
| Submand. (L+R), $D_{mean}$ | *10.5* | 10.5 | 11.2 | 11.5 | 10.9 | 10.8 |

***Table 2.*** *Table with metrics for different plans for the H&N case. Two arc plans with shoot-through are shown: uncollimated (Arc+ST) and collimated (Arc+ST+Coll). Additionally, two collimated arc plans are presented: Arc+Coll and sub-Arc+Coll. The second collimated arc consists of 2 sub-arcs and two lateral beams at 90° and 270°, with the ambition to try to reduce the air gap over the full 360-degree revolution. The number of energy layers and spots, as well as the portion delivered with shoot-through protons (% ST) are reported in the upper rows. CI and HI are reported for both CTVs (prescription levels 66 GyRBE and 55 GyRBE, respectively), as well as the worst-case value over the 28 scenarios in robust evaluation (3 mm setup and 3.5% range uncertainty) for the two CTVs as a percentage of the respective prescription dose. For the OARs, the mean doses to the External ROI, the parotids and submandibular glands are shown together with the maximum dose to the spinal cord. Both for parotids and submandibular glands the reported value is the average of the mean doses in the left and right glands. A bold value signals that the value is better than in the Arc+ST plan.*

The number of spots is notably higher for the Arc+ST and Arc+ST+Coll plans compared to the other plans. For the arc plans without ST, the number of spots is comparable to the 5-beam plans. The delivery time for the H&N case exhibits similar characteristics as the phantom plans, with a substantially lower treatment time when no collimator is used. The Arc+ST plan outperforms the collimated arc plans in terms of delivery time and is also faster than the 5-beam plans, which are both collimated. Also for the H&N case the total number of irradiated protons is lowest for the Arc+ST among all plans.

## 3.2 Range verification

Figure 5 displays the results of the simulated range verification experiment for the three error scenarios in the H&N case. The systematic 5% SPR shift shows, as expected, a consistent relative WET range difference throughout all gantry angles and spot positions. For the patient position shift in the anterior direction, the largest errors are shown for beam angles perpendicular to the position shift (90° and 270°). Most of the errors are seen in the central part of the field. For the shift of the right shoulder, errors can only be seen for a few angles for which protons pass through the shoulder (maximum error at 54° and 234°). It should be noted that the detected WET differences for 54° result from protons exiting the patient behind the target and those large range differences do not imply any deterioration of the target dose. The errors are, as expected, seen in the lower central part of the field, where the spots pass the shoulder.

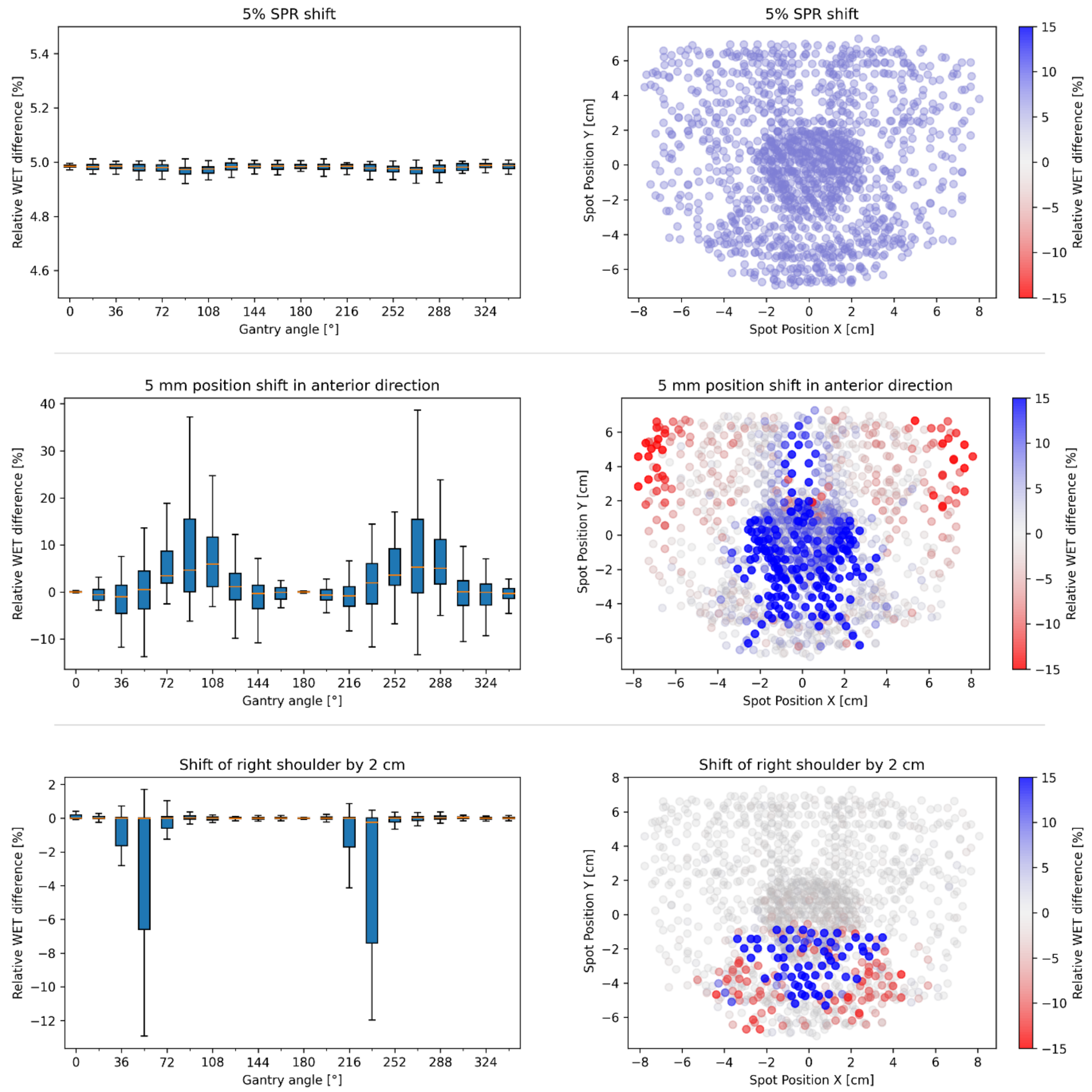

***Figure 5.*** *Detection of the relative WET differences by the ST protons for three different simulated uncertainty scenarios for the H&N patient case: 5% stopping power ratio (SPR) shift (upper panel), 5 mm position shift in anterior direction (middle panel), and 2 cm shift of right shoulder in the anterior direction. The left column shows box plots of the relative WET differences in percent as a function of the gantry angle. The orange line represents the median, the boxes show the inter-quartile range (IQR) between the first (Q1) and third quartile (Q3). The whiskers represent the highest data point within the distance Q3+1.5IQR and the lowest data point within the*

*distance Q1-1.5IQR, respectively. The right column shows the overlayed spot positions from all angles with a color scale to show where the range differences occur. Grey values represent no difference, and red (blue) values show a shorter (longer) range in the perturbed scenario. The color scale is capped at ±15%. In total, WET differences for 1422 shoot-through spots were recorded.*

# 4 Discussion

We have presented a method that combines static proton arcs in upright position with an additional ST energy layer. The study shows that the method can improve conformity and homogeneity and has the potential to remove the need for additional collimation. Layer-by-layer collimation can be an efficient way to trim individual spot dose distributions and improve plan conformity and doses to OARs[16]. However, it comes at the expense of lower dose homogeneity[16], higher neutron dose[30] and additional time in treatment plan optimization[13], as well as in treatment delivery due to mechanical movement of the adaptive aperture. Another advantage with the Arc+ST technique is that there seems to be a lower need for creating sub-arcs to reduce the air gap in different parts of the full revolution, which is beneficial in terms of both treatment planning and delivery.

Kong et al.[17] saw an increase in the number of spots in the ST plans (around 17% increase on average), but concluded that the improvements were not related to the higher number of spots. In our study, we also see increases in the number of spots for the IMPT plans with ST. This is true also for the arc plans with ST in the H&N case, but for the phantom case the number of spots is at the same level or lower for the Arc+ST compared to Arc+Coll. This confirms that the improvement with ST comes from the sharper beam penumbra rather than from an increased number of spots. The IMPT plans with ST exhibit improved plan quality indices compared to normal IMPT, but at a lower level than the Arc+ST plans, showing that the combination of ST with arcs give synergies.

The increase in number of spots and energy layers in the H&N plans for Arc+ST contributes to a longer delivery time compared to IMPT. On the other hand, the Arc+ST plans could deliver high-quality plans without a collimator, which in turn reduces treatment times compared to the collimated IMPT and arc plans. With a fast energy switching time and a rotational speed of the upright patient positioner of 1 rpm[31], the Arc+ST without adaptive aperture has an estimated delivery time of less than 5.4 minutes and in terms of delivery time outperforms the collimated arc plans, as well as the 5-beam IMPT plans. Previous studies with gantry-based proton arcs for H&N cancer patients have reported larger delivery times for a similar amount of directions: de Jong et al.[32] saw reduced toxicity in oropharyngeal cancer patients using both 20 and 30 beam directions. The average delivery time was estimated at around 11 minutes using 20 directions. In another study, de Jong et al.[33] compared static arc plans to two different algorithms for generation of dynamic arcs. The delivery time for 30-direction static arc plans was estimated at around 10 minutes and around 8 minutes for the dynamic arc plans with 360 energy layers. Engwall et al.[14] proposed to partition static proton arcs with in total 30 directions into subplans with fewer directions to be delivered over different fractions in order to reduce delivery time while still utilizing many of the dosimetric benefits of proton arcs. Static arcs with 30 directions showed a delivery time of just above 10 minutes and when reducing it to 10 directions per fraction the delivery time was around 5 minutes. The clinically implemented static arc treatments reported by Frachiolla et al.[12] are delivered as multiple IMPT beams, which introduces additional delays between the beams and the average measured delivery time for H&N plans with 20 beams is around 25 minutes. Our estimated delivery times compare well with these previously published results. However, a fair comparison between the different techniques should include both delivery times and plan quality, which is outside the scope of this paper. Furthermore, the time model is based on rough estimates, as there are no published measurements of the delivery parameters—unlike other delivery systems for which such data are available[34–36]. The timings should therefore be verified in a real delivery setting.

In this study the percentage of ST in the arc plans was between 24% to 45% of the total number of irradiated protons when no collimator was used. Since the ST protons have their Bragg peak outside the patient, the relative portion of the total dose delivered by ST protons will be lower than the reported percentage of irradiated protons, but they will still contribute to the total dose as can be seen in Figures 2 and 4. The optimizer selects freely where to employ the ST protons and at the end of the optimization they will be distributed dependent on the dose-based objective functions. To increase conformity, a natural placement of the ST contribution is at the border of the target (see Figure 2, Figure 4 and Figure S3). Since the ST protons will traverse the full patient geometry, the mean dose to the External will see an increase. However, for a complex geometry like H&N the increase only seems to be marginal. This initial finding should be confirmed in a larger patient cohort. Another interesting finding is that the total number of irradiated protons is substantially lower in the Arc+ST plans than in any of the collimated plans in this study. This means that a large portion of the delivered protons in the collimated plans are stopped in the aperture, which leads to activation in the aperture and generation of secondary neutrons. Since a significant portion of the dose in the Arc+ST plans is delivered without any range shifter plates in the beam line, the secondary neutron production is further reduced.

For the IMPT plans, the percentage of ST protons was substantially lower than the corresponding Arc plans. This is expected based on the common optimization problem formulation for IMPT and Arc: the dose fall-off functions suppress the ST layers to a higher extent in IMPT, where the exit dose is not distributed over a large number of directions. The observed behavior confirms the hypothesis that the combination of ST with proton arcs can be more favorable than in IMPT. We also note that the percentage of ST protons for the IMPT case is lower than in Kong et al.[17] (on average 28% for oropharyngeal cancer patients compared to 16% for the IMPT plan for the H&N case in our study). This difference could mainly be attributed to the use of different treatment machines, the use of a collimator for the IMPT plans in our study, a larger number of beam directions in Kong et al.[17] (6 vs 5), as well as a higher ST energy (244 vs 230 MeV), and different optimization problem formulations.

Previous studies have shown that PAT could result in improved LET distributions[37–39] by moving high-LET end-of-range protons from OARs, which in turn leads to a more beneficial radiobiology[40]. The ST protons will per construction deliver the high-LET outside the patient and thus not in the OARs. The LET distribution could potentially be further improved by combining Arc+ST with LET optimization or optimization with a variable RBE model based on LET. This is a topic that warrants future research.

In addition to the Arc+ST plans, we also created uncollimated static arc plans with only ST protons (Full-ST) for the phantom targets. These plans were inferior to the Arc+ST plans, especially for the larger targets (see Figure S2 in the supplementary material). This is not surprising since this mode only delivers protons from a discrete number of directions (20 directions in this study). Full-ST in the dynamic arc mode is likely to yield better results by utilizing more delivery angles and could serve as an alternative to photon VMAT treatments at existing proton facilities. Reports have seen improvements with Full-ST IMPT in stereotactic cases in lung compared to conformal static photon beams[41] and VMAT[42], as well as in H&N cases compared to VMAT[43]. For the CTV Small in Figure S2, we see that the conformity, homogeneity, and robustness are on the same level for Full-ST as for Arc+ST, but at the expense of a higher mean dose to the External. Advantages of employing only transmission protons could include an even sharper lateral penumbra combined with reduced concerns about range robustness and high-LET effects at the end of range[44]. In dynamic mode, Full-ST arcs are simpler to deliver than Bragg-peak (BP) arcs, which in many cases require complex sequences of energy switches. The Arc+ST plans in this study are mostly suited for delivery in static mode since several energy layers are already delivered from each discrete direction. However, it would still be feasible to create similar plans using a dynamic arc approach, but that would require either two revolutions (one with the ST portion and one with the BP portion) or incorporating the ST layers in the energy layer sequencing of dynamic arcs. An alternative hybrid approach could include static directions with dynamic ST segments between them.

While the concept with static ST proton arcs could be delivered with a gantry, upright arcs come with the additional advantage of a single beam dump opposite the fixed beam line and possibilities for instantaneous and simple range verification from a multitude of directions using a fixed range detector. Methods already exist to provide sufficient time resolution to determine individual ST proton ranges behind the patient by multi-layer ionization chambers and they have been implemented in a clinical setting[25–27]. Table 3 shows a comparison of our proposed method for range verification for upright treatments versus these existing methods. Some of the major advantages with our method include the ability to perform range verification for all planning angles and not only one horizontal angle. This makes our method applicable for any proton treatment technique in upright position, regardless of treatment angles, and hence easier to adopt in clinical practice. Furthermore, our method avoids making patient setup slower because the range detector is always present in the treatment room and the ST protons are an integral part of the treatment plan. As such, the method can be used during treatment in every treatment fraction and is therefore sensitive to both intra- and inter-fractional uncertainties. One limitation of the current approach is that we use a ray-trace method, which could have insufficient accuracy, especially in heterogeneous regions. If this proof-of-concept is to be translated to a clinical setting, Monte Carlo methods should be considered.

A topic of future research studies, could be to investigate the combination of the range verification method with  techniques developed for proton tomography[24].The large number of directions in PAT plan would allow for making a back-projection of the range differences to reconstruct where in the patient 3D geometry the uncertainties arise. For the H&N case in this study, the total number of ST spots were 1422, which is not sufficient to create a high-resolution proton CT, but they could potentially provide low-resolution information where in the proton beam paths the errors occur. This low-resolution image could be overlayed on the planning CT, a repeat CT or CBCT to get better spatial information.  Another interesting use case to explore is a setting where the range verification setup  is used as a trigger to interrupt a treatment in real time if the range errors exceed some pre-defined limit.

| | **Proposed method** | **Previous work[26,27]** |
|---|---|---|
| ***Treatment technique*** | Upright with fixed range detector | Gantry-based with portable range detector |
| ***Angles used for range verification*** | All angles used in the plan | Only horizontal |
| ***Increased time for patient setup*** | No | Yes |
| ***Contributes to target conformity*** | Yes | No |
| ***Dose*** | High. Contributes to target coverage. | ~1 cGy |
| ***Improves plan quality*** | Yes | No |
| ***Sensitive to intra- and inter-fractional changes*** | Yes.<br>The ST protons are an integral part of the plan and will be used in every fraction and during delivery | • *Intra-fractional:* No. The range verification is performed prior to treatment.<br>• *Inter-fractional:* Possibly, if used in every fraction, despite the increased time for patient setup |

***Table 3.*** *Comparison of the proposed upright range verification method with Arc+ST to previously presented gantry-based range verification methods[26,27].*

In this paper, we have provided a proof-of-concept of ST arcs in upright position. More work remains to take the concept to actual delivery and to develop the method for accurate range probing. Future studies should also include more patient cases in order to make a thorough statistical analysis of the potential improvements with the new technique.

# 5 Conclusion

Combination of static proton arcs with ST layers leads to dosimetric advantages and reduces the need for collimation in compact proton delivery systems with energy selection performed close to the patient. The removal of collimation leads to a large reduction of the delivery time and the Arc+ST plan for the H&N case is estimated to be delivered within 5.4 minutes, which is faster than all other studied plans for this case, including 5-beam IMPT. When delivered in upright position, the ST protons can by simple means be used for instantaneous range verification. Given the proof-of-concept nature of this study, we foresee that more work is needed in terms of clinical implementation and validation in a large patient cohort. Nevertheless, the study has highlighted combined advantages of the new treatment technique, which could leverage the potential of compact upright proton treatments and thus make proton treatments more affordable and accessible to a larger patient group.

## Acknowledgements

The authors would like to thank Axel Grönlund for his design work for Figure 1.

## Conflict of Interest Statement

All authors are employees of RaySearch Laboratories.

1. Mein S, Wuyckens S, Li X, et al. Particle arc therapy: Status and potential. *Radiother Oncol*. 2024;199:110434. doi:10.1016/j.radonc.2024.110434

2. De Jong BA, Battinelli C, Free J, et al. Spot scanning proton arc therapy reduces toxicity in oropharyngeal cancer patients. *Med Phys*. 2023;50(3):1305-1317.

3. de Jong BA, Korevaar EW, Maring A, et al. Proton arc therapy increases the benefit of proton therapy for oropharyngeal cancer patients in the model based clinic. *Radiother Oncol*. 2023;184:109670. doi:10.1016/j.radonc.2023.109670

4. de Jong BA BS Maring A, Werkman CI, Cannavo D, van der Schaaf A, Scandurra D, Engwall E, Korevaar EW, Janssens G, Langendijk JA. Toxicity benefit and inter-fraction robustness of proton arc therapy compared to IMPT and VMAT for nasopharyngeal cancer patients. *Med Phys*. Published online 2023.

5. Ding X, Li X, Qin A, et al. Have we reached proton beam therapy dosimetric limitations? – A novel robust, delivery-efficient and continuous spot-scanning proton arc (SPArc) therapy is to improve the dosimetric outcome in treating prostate cancer. *Acta Oncol*. 2018;57(3):435-437. doi:10.1080/0284186X.2017.1358463

6. Li X, Kabolizadeh P, Yan D, et al. Improve dosimetric outcome in stage III non-small-cell lung cancer treatment using spot-scanning proton arc (SPArc) therapy. *Radiat Oncol*. 2018;13(1):35. doi:10.1186/s13014-018-0981-6

7. Chang S, Liu G, Zhao L, et al. Feasibility study: spot-scanning proton arc therapy (SPArc) for left-sided whole breast radiotherapy. *Radiat Oncol*. 2020;15(1):232. doi:10.1186/s13014-020-01676-3

8. Ding X, Zhou J, Li X, et al. Improving dosimetric outcome for hippocampus and cochlea sparing whole brain radiotherapy using spot-scanning proton arc therapy. *Acta Oncol*. 2019;58(4):483-490. doi:10.1080/0284186X.2018.1555374

9. Liu G, Li X, Qin A, et al. Improve the dosimetric outcome in bilateral head and neck cancer (HNC) treatment using spot-scanning proton arc (SPArc) therapy: a feasibility study. *Radiat Oncol*. 2020;15(1):21. doi:10.1186/s13014-020-1476-9

10. Liu G, Li X, Qin A, et al. Is proton beam therapy ready for single fraction spine SBRS? – a feasibility study to use spot-scanning proton arc (SPArc) therapy to improve the robustness and dosimetric plan quality. *Acta Oncol*. 2021;60(5):653-657. doi:10.1080/0284186X.2021.1892183

11. Liu G, Zhao L, Qin A, et al. Lung Stereotactic Body Radiotherapy (SBRT) Using Spot-Scanning Proton Arc (SPArc) Therapy: A Feasibility Study. *Front Oncol*. 2021;11:664455. doi:10.3389/fonc.2021.664455

12. Fracchiolla F, Engwall E, Mikhalev V, et al. Static proton arc therapy: Comprehensive plan quality evaluation and first clinical treatments in patients with complex head and neck targets. *Med Phys*. Published online February 12, 2025:mp.17669. doi:10.1002/mp.17669

13. Janson M, Glimelius L, Fredriksson A, Traneus E, Engwall E. Treatment planning of scanned proton beams in RayStation. *Med Dosim*. 2024;49(1):2-12. doi:10.1016/j.meddos.2023.10.009

14. Engwall E, Marthin O, Wase V, et al. Partitioning of discrete proton arcs into interlaced subplans can bring proton arc advances to existing proton facilities. *Med Phys*. 2023;50(9):5723-5733.

15. Feldman J, Pryanichnikov A, Achkienasi A, et al. Commissioning of a novel gantry-less proton therapy system. *Front Oncol*. 2024;14:1417393. doi:10.3389/fonc.2024.1417393

16. Silvus A, Haefner J, Altman MB, Zhao T, Perkins S, Zhang T. Dosimetric evaluation of dose shaping by adaptive aperture and its impact on plan quality. *Med Dosim*. 2024;49(1):30-36. doi:10.1016/j.meddos.2023.10.011

17. Kong W, Huiskes M, Habraken SJM, et al. Reducing the lateral dose penumbra in IMPT by incorporating transmission pencil beams. *Radiother Oncol*. 2024;198:110388. doi:10.1016/j.radonc.2024.110388

18. Amstutz F, Krcek R, Bachtiary B, et al. Treatment planning comparison for head and neck cancer between photon, proton, and combined proton–photon therapy – From a fixed beam line to an arc. *Radiother Oncol*. 2024;190:109973. doi:10.1016/j.radonc.2023.109973

19. Parodi K, Polf JC. *In vivo* range verification in particle therapy. *Med Phys*. 2018;45(11). doi:10.1002/mp.12960

20. Verburg JM, Seco J. Proton range verification through prompt gamma-ray spectroscopy. *Phys Med Biol*. 2014;59(23):7089-7106. doi:10.1088/0031-9155/59/23/7089

21. Berthold J, Pietsch J, Piplack N, et al. Detectability of Anatomical Changes With Prompt-Gamma Imaging: First Systematic Evaluation of Clinical Application During Prostate-Cancer Proton Therapy. *Int J Radiat Oncol*. 2023;117(3):718-729. doi:10.1016/j.ijrobp.2023.05.002

22. Parodi K, Enghardt W, Haberer T. In-beam PET measurements of $\beta^+$ radioactivity induced by proton beams. *Phys Med Biol*. 2002;47(1):21-36. doi:10.1088/0031-9155/47/1/302

23. Ozoemelam I, Van Der Graaf E, Van Goethem MJ, et al. Feasibility of quasi-prompt PET-based range verification in proton therapy. *Phys Med Biol*. 2020;65(24):245013. doi:10.1088/1361-6560/aba504

24. Poludniowski G, Allinson NM, Evans PM. Proton radiography and tomography with application to proton therapy. *Br J Radiol*. 2015;88(1053):20150134. doi:10.1259/bjr.20150134

25. Farace P, Righetto R, Meijers A. Pencil beam proton radiography using a multilayer ionization chamber. *Phys Med Biol*. 2016;61(11):4078-4087. doi:10.1088/0031-9155/61/11/4078

26. Meijers A, Seller Oria C, Free J, Langendijk JA, Knopf AC, Both S. Technical Note: First report on an in vivo range probing quality control procedure for scanned proton beam therapy in head and neck cancer patients. *Med Phys*. 2021;48(3):1372-1380. doi:10.1002/mp.14713

27. Seller Oria C, Thummerer A, Free J, et al. Range probing as a quality control tool for CBCT-based synthetic CTs: In vivo application for head and neck cancer patients. *Med Phys*. 2021;48(8):4498-4505. doi:10.1002/mp.15020

28. Paddick I. A simple scoring ratio to index the conformity of radiosurgical treatment plans: Technical note. *J Neurosurg*. 2000;93(supplement_3):219-222. doi:10.3171/jns.2000.93.supplement_3.0219

29. Mumot M, Algranati C, Hartmann M, Schippers JM, Hug E, Lomax AJ. Proton range verification using a range probe: definition of concept and initial analysis. *Phys Med Biol*. 2010;55(16):4771-4782. doi:10.1088/0031-9155/55/16/010

30. Smith BR, Hyer DE, Hill PM, Culberson WS. Secondary Neutron Dose From a Dynamic Collimation System During Intracranial Pencil Beam Scanning Proton Therapy: A Monte Carlo Investigation. *Int J Radiat Oncol*. 2019;103(1):241-250. doi:10.1016/j.ijrobp.2018.08.012

31. Boisbouvier S, Boucaud A, Tanguy R, Grégoire V. Upright patient positioning for pelvic radiotherapy treatments. *Tech Innov Patient Support Radiat Oncol*. 2022;24:124-130. doi:10.1016/j.tipsro.2022.11.003

32. de Jong BA, Battinelli C, Free J, et al. Spot scanning proton arc therapy reduces toxicity in oropharyngeal cancer patients. *Med Phys*. Published online October 2022. doi:https://doi.org/10.1002/mp.16098

33. de Jong BA, Zhao L, Liu P, et al. Early generation dynamic and static proton arc treatment planning algorithms assessment in oropharyngeal cancer patients. *Med Phys*. Published online 2025. doi:10.1002/mp.17916

34. Pfeiler T, Bäumer C, Engwall E, Geismar D, Spaan B, Timmermann B. Experimental validation of a 4D dose calculation routine for pencil beam scanning proton therapy. *Z Für Med Phys*. 2018;28(2):121-133. doi:10.1016/j.zemedi.2017.07.005

35. Zhao L, Liu G, Zheng W, et al. Building a precise machine-specific time structure of the spot and energy delivery model for a cyclotron-based proton therapy system. *Phys Med Biol*. 2022;67(1):01NT01. doi:10.1088/1361-6560/ac431c

36. Shen J, Tryggestad E, Younkin JE, et al. Technical Note: Using experimentally determined proton spot scanning timing parameters to accurately model beam delivery time. *Med Phys*. 2017;44(10):5081-5088. doi:10.1002/mp.12504

37. Bertolet A, Carabe A. Proton monoenergetic arc therapy (PMAT) to enhance LETd within the target. *Phys Med Biol*. 2020;65(16):165006. doi:10.1088/1361-6560/ab9455

38. Li X, Ding X, Zheng W, et al. Linear Energy Transfer Incorporated Spot-Scanning Proton Arc Therapy Optimization: A Feasibility Study. *Front Oncol*. 2021;11:698537. doi:10.3389/fonc.2021.698537

39. Glimelius L, Marthin O, Wase V, et al. Robust LET optimization of proton arcs can substantially reduce high LET in critical structures. In: Vol 2023.

40. Carabe A, Karagounis IV, Huynh K, et al. Radiobiological effectiveness difference of proton arc beams versus conventional proton and photon beams. *Phys Med Biol*. 2020;65(16):165002. doi:10.1088/1361-6560/ab9370

41. Mou B, Beltran CJ, Park SS, Olivier KR, Furutani KM. Feasibility of Proton Transmission-Beam Stereotactic Ablative Radiotherapy versus Photon Stereotactic Ablative Radiotherapy for Lung Tumors: A Dosimetric and Feasibility Study. *PLOS ONE*. 2014;9(6):e98621. doi:10.1371/journal.pone.0098621

42. Van Marlen P, Dahele M, Folkerts M, Abel E, Slotman BJ, Verbakel WFAR. Bringing FLASH to the Clinic: Treatment Planning Considerations for Ultrahigh Dose-Rate Proton Beams. *Int J Radiat Oncol*. 2020;106(3):621-629. doi:10.1016/j.ijrobp.2019.11.011

43. Van Marlen P, Dahele M, Folkerts M, Abel E, Slotman BJ, Verbakel W. Ultra-High Dose Rate Transmission Beam Proton Therapy for Conventionally Fractionated Head and Neck Cancer: Treatment Planning and Dose Rate Distributions. *Cancers*. 2021;13(8):1859. doi:10.3390/cancers13081859

44. Verhaegen F, Wanders RG, Wolfs C, Eekers D. Considerations for shoot-through FLASH proton therapy. *Phys Med Biol*. 2021;66(6):06NT01. doi:10.1088/1361-6560/abe55a

## Supplementary material

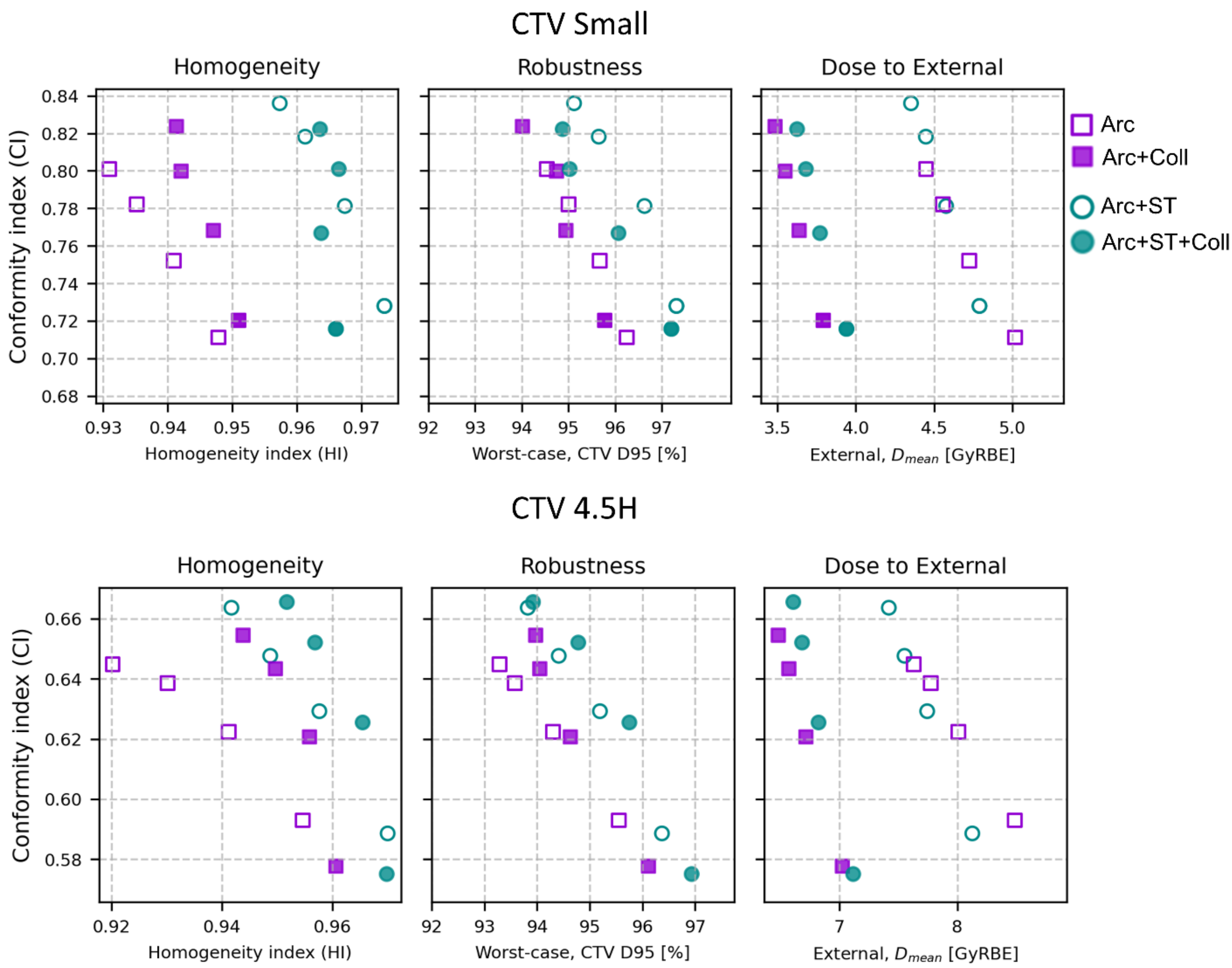

**Figure S1.** Same evaluation as in Figure 3 but for CTV Small and CTV 4.5H.

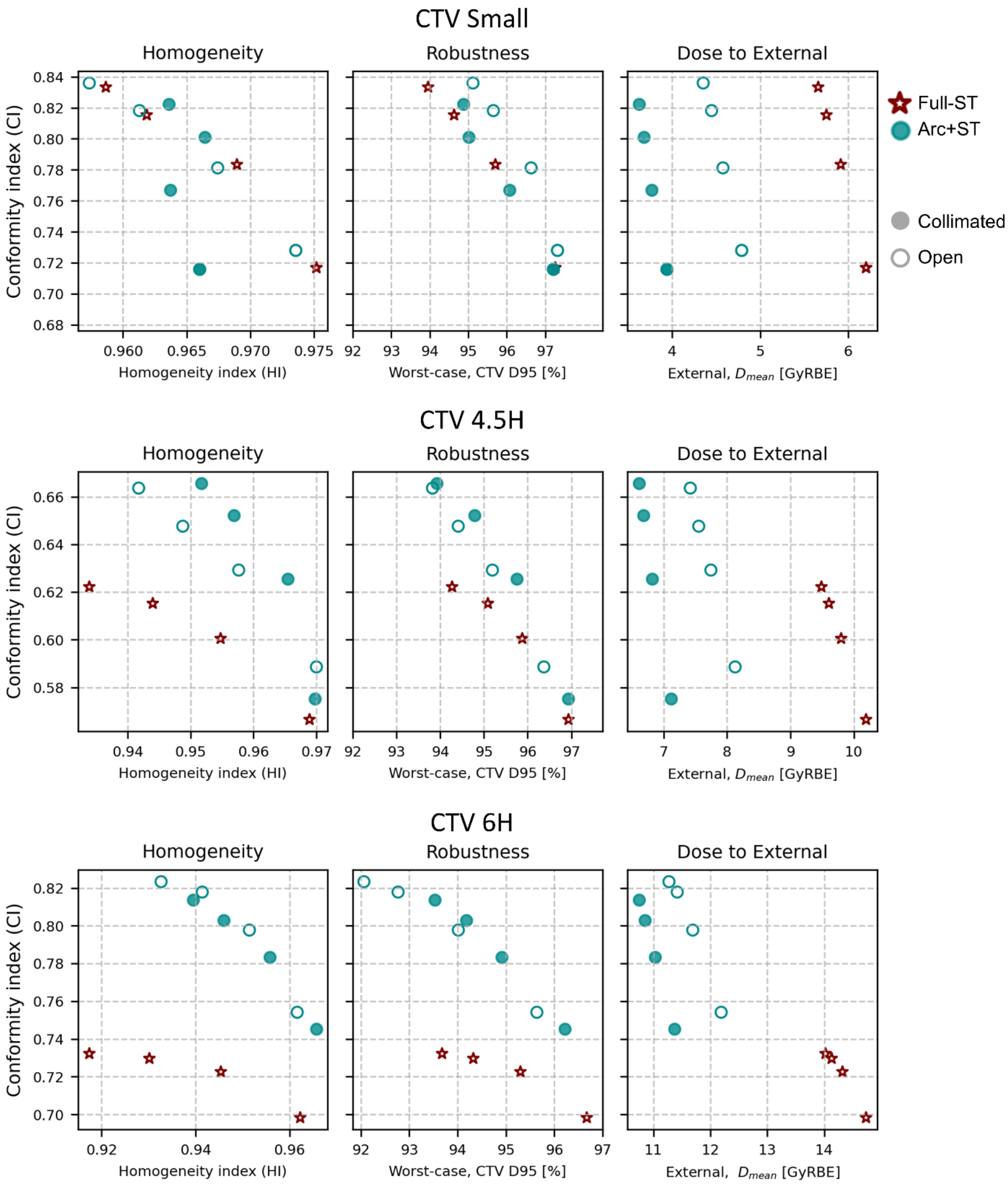

**Figure S2.** Comparison between Arc+ST and Full-ST for CTV Small, CTV 4.5H and CTV 6H. The Full-ST plans have the same number of directions (20) as the Arc+ST, but only employs a single ST layer from each direction.

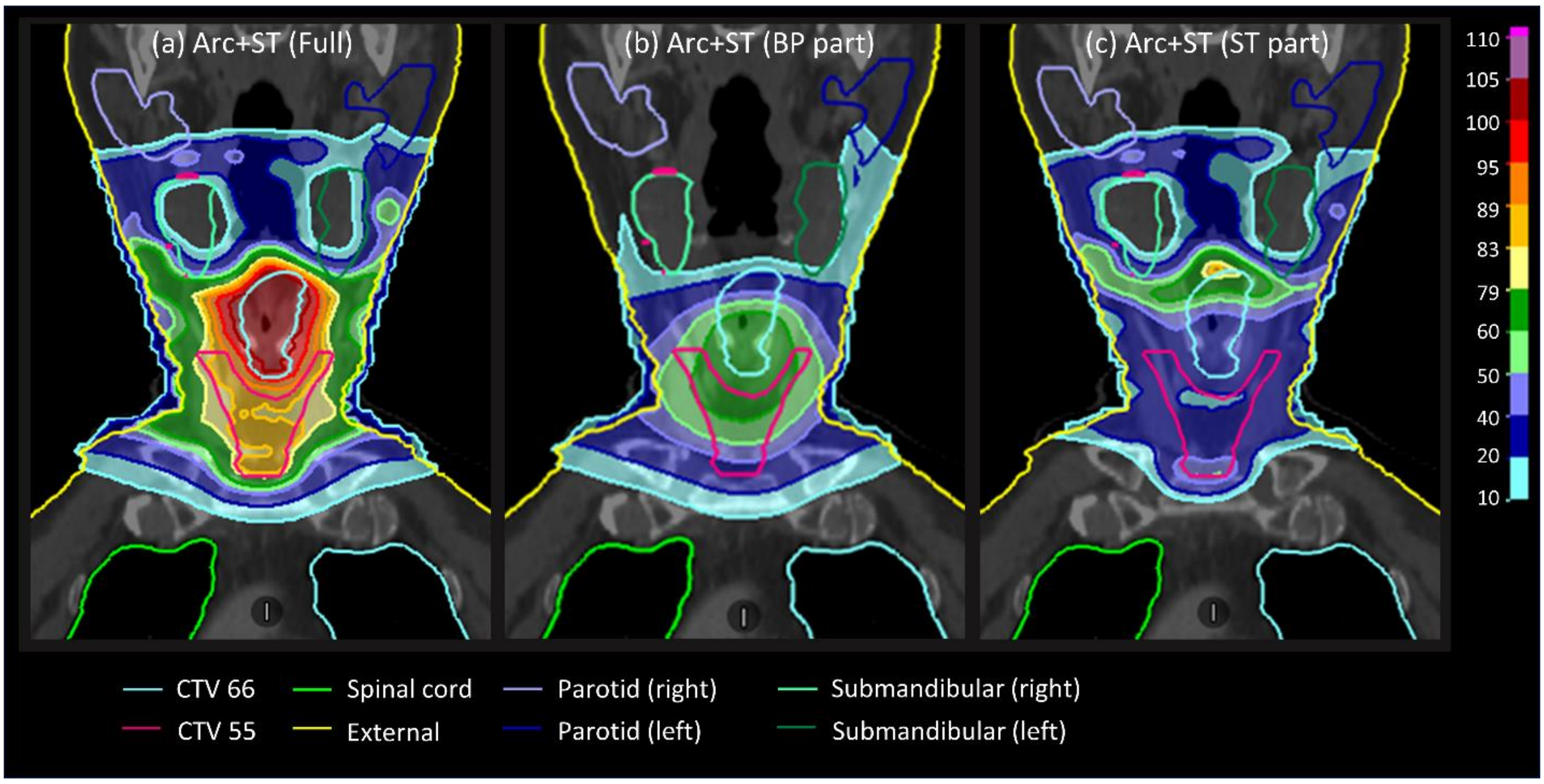

**Figure S3.** Coronal view of the Arc+ST plan. Panel (a) shows the full plan, whereas panels (b) and (c) show the Bragg peak (BP) and ST parts, respectively. It can be seen that a large portion of the ST protons are placed at the border between the CTV 66 and the submandibular glands.

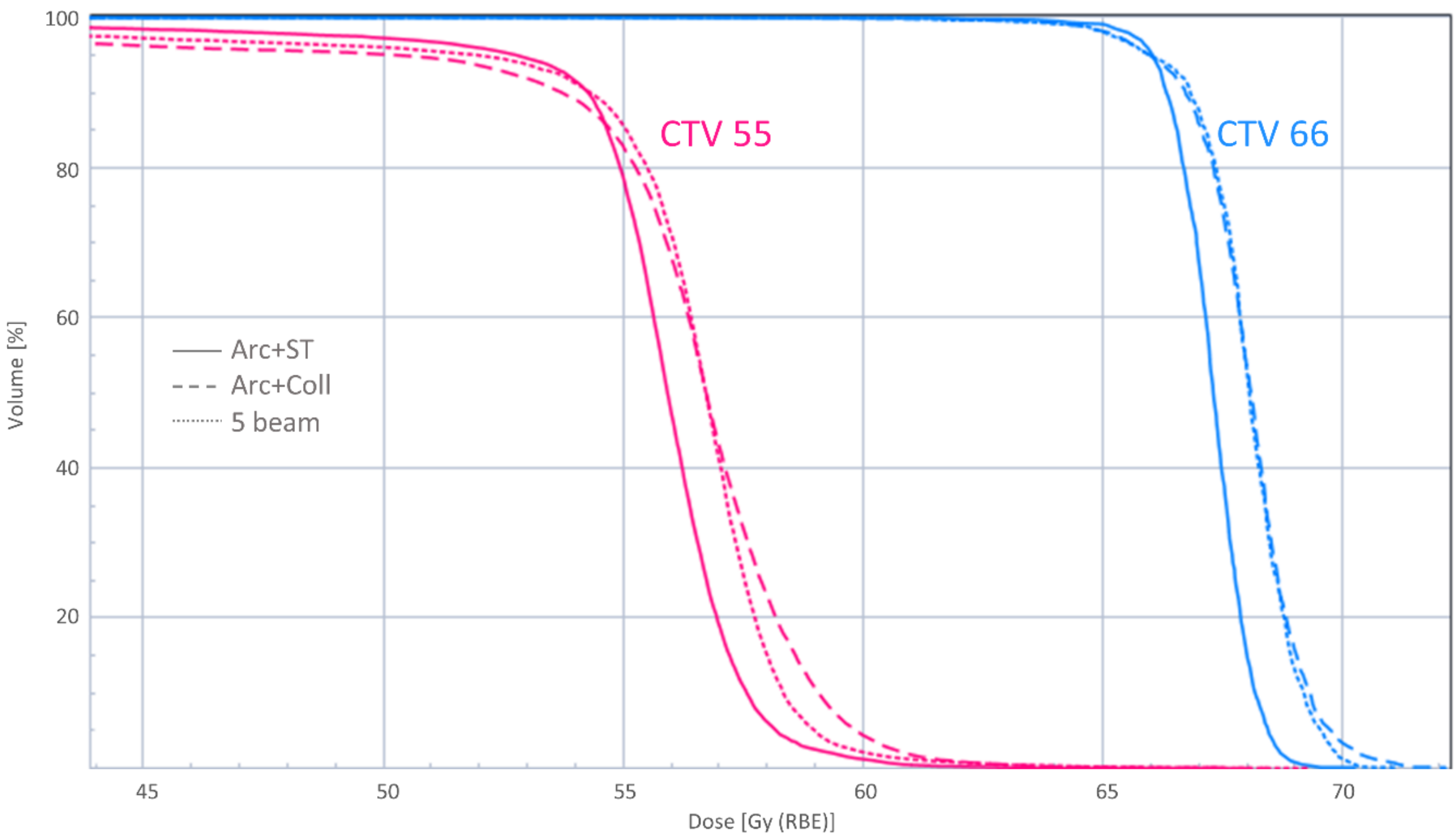

**Figure S4.** Enlargement of the DVHs in Figure 4 for the two targets CTV 55 (pink) and CTV 66 (blue) for Arc+ST (solid), Arc+Coll (dashed) and 5-beam (dotted) plans.

| | Arc+ST | sub-Arc+ST | sub-Arc |
|---|---|---|---|
| No. EL | *360* | 346 | 360 |
| No. spots | *4281* | 3797 | 3507 |
| % ST | *44.84* | 41.41 | 0 |
| CI, CTV66 | *0.48* | 0.48 | 0.42 |
| CI, CTV55 | *0.36* | 0.36 | 0.33 |
| HI, CTV66 | *0.96* | **0.97** | 0.95 |
| HI, CTV55 | *0.91* | 0.91 | 0.83 |
| Worst-case, CTV66 D95 [%] | *99.2%* | 99.2% | 98.2% |
| Worst-case, CTV55 D95 [%] | *76.5%* | **77.1%** | 70.3% |
| External, $D_{mean}$ [$Gy_{RBE}$] | *6.4* | **6.2** | 7.1 |
| SpinalCord, $D_{max}$ [$Gy_{RBE}$] | *20.1* | 21.9 | 23.3 |
| Parotid (L+R), $D_{mean}$ [$Gy_{RBE}$] | *10.3* | 10.2 | 10.6 |
| Submandibular (L+R), $D_{mean}$ [$Gy_{RBE}$] | 10.7 | 10.7 | 11.6 |

**Table S1.** Complementary table to Table 2 with more plans for the H&N case. The Arc+ST plan is included also here as the reference. The sub-Arc plans are uncollimated with and without ST.